\documentclass{INTERSPEECH2023}
\usepackage{multicol}
\usepackage{multirow}
\usepackage{enumitem}
\usepackage{booktabs}
\usepackage{amssymb}
\interspeechcameraready

\title{Detection of Cross-Dataset Fake Audio Based on Prosodic and Pronunciation Features}
\name{Chenglong Wang$^{1,2}$, Jiangyan Yi$^{2,3}$, Jianhua Tao$^{3,4}$, Chu Yuan Zhang$^{2,3}$, Shuai Zhang$^{4}$, Xun Chen$^{1}$}
\address{
	$^1$University of Science and Technology of China, Hefei, China\\
	$^2$State Key Laboratory of Multimodal Artificial Intelligence Systems, Institute of Automation, Chinese Academy of Sciences, Beijng, China \\
	$^3$School of Artificial Intelligence, University of Chinese Academy of Sciences, China \\
	$^4$Department of Automation, Tsinghua University}
\email{chenglong.wang@nlpr.ia.ac.cn}

\begin{document}

\maketitle
 
\begin{abstract}
% 1000 characters. ASCII characters only. No citations.
Existing fake audio detection systems perform well in in-domain testing, but still face many challenges in out-of-domain testing. This is due to the mismatch between the training and test data, as well as the poor generalizability of features extracted from limited views. To address this, we propose multi-view features for fake audio detection, which aim to capture more generalized features from prosodic, pronunciation, and wav2vec dimensions. Specifically, the phoneme duration features are extracted from a pre-trained model based on a large amount of speech data. For the pronunciation features, a Conformer-based phoneme recognition model is first trained, keeping the acoustic encoder part as a deeply embedded feature extractor. Furthermore, the prosodic and pronunciation features are fused with wav2vec features based on an attention mechanism to improve the generalization of fake audio detection models. Results show that the proposed approach achieves significant performance gains in several cross-dataset experiments.
\end{abstract}
\noindent\textbf{Index Terms}: fake audio detection, ASVspoof, prosodic feature, pronunciation feature, cross-dataset

\section{Introduction}

Currently, there are several types of front-end features used for fake audio detection. These include short-time spectral features, raw audio, fundamental frequency features, and self-supervised features \cite{witkowski2017audio, font2017experimental, novoselov2016stc, korshunov2016cross, das2019long}.  Todisco\cite{todisco2017constant} demonstrated the superior performance of Constant Q Cepstral Coefficients (CQCC) over Mel Frequency Cepstral Coefficients (MFCC) by using the constant Q transform to process the speech signal. Sahidullah\cite{sahidullah2015comparison} proposed Linear Frequency Cepstrum Coefficients (LFCC) by replacing the Mel scale filters with linear filters, which focuses more on high frequency band features compared to MFCC. Das\cite{das2020assessing} improved CQCC features by proposing Extended Constant-Q Cepstral Coeffficients (eCQCC) features and Constant-Q Statistics-Plus-Principal Information Coefficient (CQSPIC) features. These methods have shown promising results in in-domain tests, achieving equal error rates (EERs) of less than 1\% in the ASVspoof2019 logical access (LA) scenarios \cite{todisco2019asvspoof}. In addition, Tak et al.\cite{tak2021end} advocate for using raw audio directly as input, as they believe that front-end features relying on empirical design lose much of the speech information. 

For prosodic features, fundamental frequency has been investigated by several scholars \cite{xiao2015spoofing, attorresi2022combining, kamble2020advances, patil2017novel}. Prosodic features are extracted from longer speech segments, such as phonemes and syllables, to capture the style and intonation of speech \cite{wu2015spoofing}. Patel\cite{patel2016effectiveness} improved the detection of fake audio considerably by fusing F0 contour and 36-D MFCC at score level. Pal\cite{pal2018synthetic} proposed fundamental frequency variation features to capture the prosodic difference between real and fake audio. Xue\cite{xue2022audio} proposed the F0 subband, which fully utilizes F0 information and provides a new and effective basis for frequency band division. Another popular feature extractor for fake audio detection is wav2vec, which is based on the self-supervised method \cite{lv2022fake, wang2021investigating, martin2022vicomtech}. The wav2vec features obtained after training with a large amount of unlabeled data have achieved first place in many competitions \cite{lv2022fake, martin2022vicomtech}.

\begin{figure*}[t]
	\centering
	
	\includegraphics[height=7cm,width=0.85\textwidth]{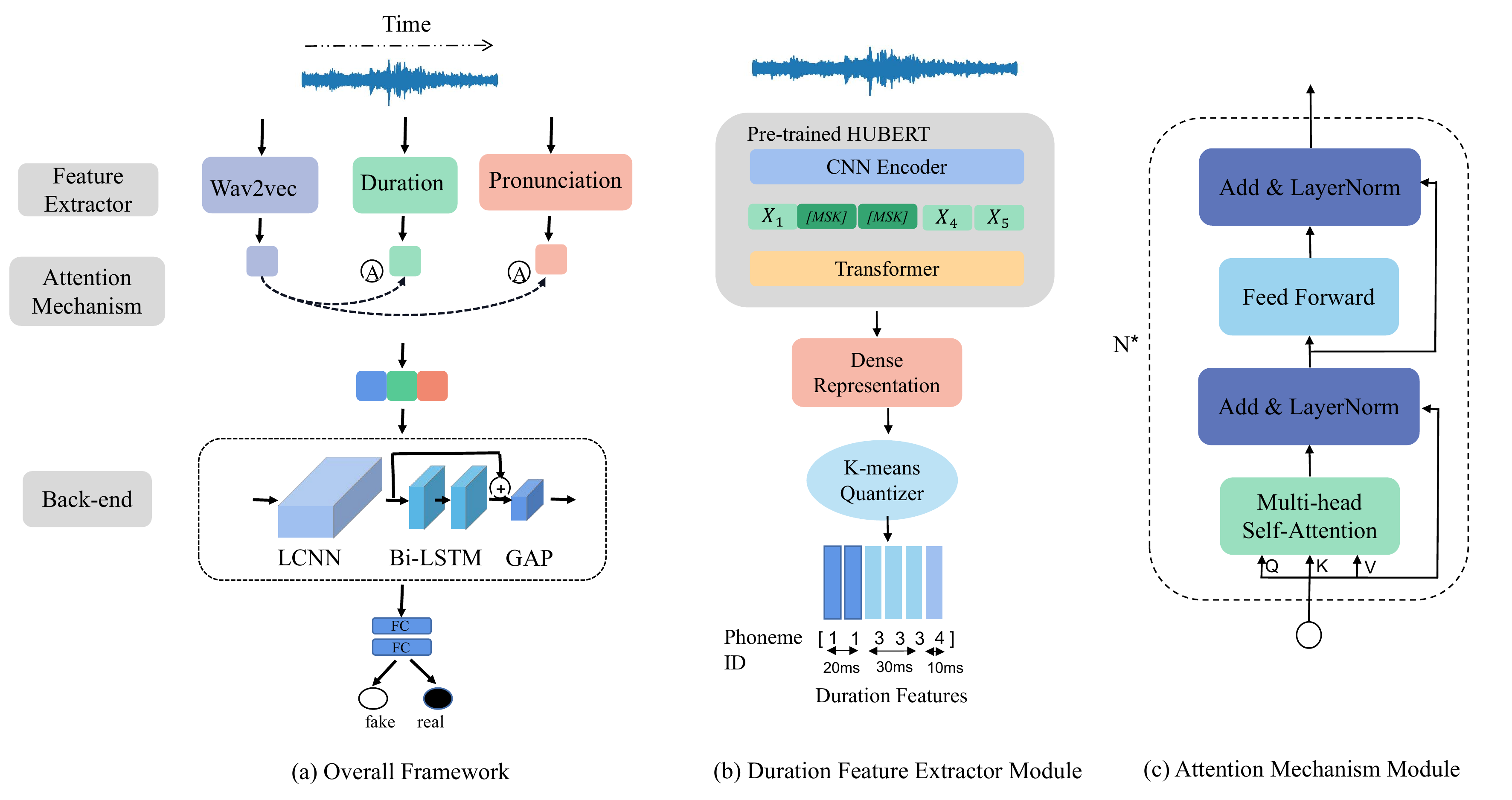}
	\caption{
		(a) Overall framework of proposed method. The system consists of the feature extraction module, LLGF block, and the attention mechanism module. \textcircled{A} denotes the attention mechanism. (b) Duration Encoder Module. The phoneme ID "134" in the figure refers to three different phonemes respectively. (c) The attention mechanism module.
	}
	
	\label{fig:speech_production}
\end{figure*}

In practical applications, detecting fake audio can be challenging due to the poor generalization of existing systems to unknown types of spoofing attacks. This limitation is mainly attributed to the fact that features extracted from a single dimension often lack generalization capability. For instance, short-time spectral features, which are extracted from short frames of 20-30 ms, are vulnerable to channel effects. In addition, current methods for prosodic feature extraction only consider F0 features and overlook critical phoneme duration features. Notably, the duration of the same phoneme may vary significantly in different real audio contexts, while the duration of phonemes in fake audio is usually more uniform. Furthermore, it is difficult to identify which speech information is contained in the self-supervised features generated by wav2vec. Therefore, we propose the use of multi-view features for enhancing the detection of fake audio, which can improve the generalization of detection across datasets. Our approach incorporates features from three dimensions: prosodic, pronunciation, and wav2vec. We present phoneme duration extractors and pronunciation feature extractors to achieve this. To obtain phoneme-like duration features, we encode the speech using the pre-training model HuBERT\cite{hsu2021hubert} without the transcript. For the pronunciation features, we first train a Conformer-based \cite{gulati2020conformer} phoneme recognition model. Then keep the acoustic encoder part as a deeply embedded feature extractor. We further fuse the prosodic and pronunciation features with discrete clustering-based wav2vec features through an attention mechanism to improve the performance of fake audio detection. Our results show that these auxiliary features can be used to improve the detection performance both in and out of the domain. The main contributions of this study can be summarized as follows:
\begin{itemize}
	\item We propose pronunciation features and phoneme duration features for fake audio detection for the first time.
	\item We use the attention mechanism approach to effectively fuse the prosodic features and pronunciation features with wav2vec features.
	
\end{itemize}

The rest of this paper is organized as follows: Section 2 illustrates our method. Experiments, results and discussions are reported in Section 3 and 4, respectively. Finally, we conclude the paper in Section 5.
\section{Our Method}
Our model consists of three major modules. Firstly, we used the feature extractor to extract three types of features, including wav2vec features, phoneme duration features, and pronunciation features. It should be noted that we believe the duration of phonemes can reflect prosodic information, therefore in this paper, the term “prosodic feature” refers to the duration of phonemes. Second, the attention module fuses prosodic and pronunciation features with different weights to the wav2vec features. Finally, the back-end learns a deep representation of speech. The architecture of our approach is shown in Figure 1 (a), and we provide further details in the following sections.

\subsection{Features}
\textbf{Duration:} Since existing publicly available fake audio datasets such as ASVspoof, ADD2022, etc., do not provide audio-corresponding transcript, we cannot extract the duration information of the fake audio by forcing the alignment. Inspired by \cite{kharitonov2022textless}, we encode the speech with the pre-training model HuBERT, and the resulting encoding vector is an encoding similar to speech phonemes. As shown in Figure 1 (b), the first step is to encode the original audio into an encoding vector with a HuBERT model which is pre-trained on the LibriSpeech corpus. We choose k-means as the quantization operation to transform the output of the encoder from continuous to discrete values. Formally, ${D_i} = Q({C_i})$, where $Q$ is the quantization function k-means, ${C_i}$ is a sequence of vectors, ${D_i} = [{d_1},{d_2},{\ldots}, {d_T}]$ such that ${d_i} \in \{1,2,{\ldots},K\}$ and $K$ is the size of the phoneme vocabulary, we set $K=100$. We refer to the final obtained ${D_i}$(e.g.,$[1,1,3,3,3,4]$) as the phoneme duration vector.

\begin{figure}[tb]

	\centering
	\centerline{\includegraphics[width=9cm, height=2.6cm]{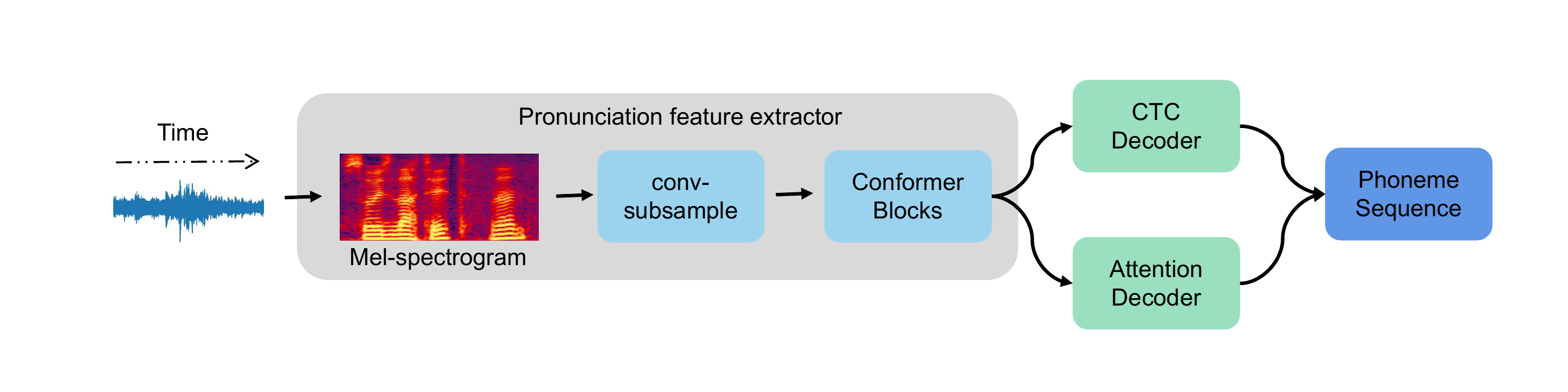}}
	%  \vspace{2.0cm}
	%\centerline{(a) Original audio}\medskip

	\caption{Pronunciation Feature Extractor Module}
	\label{fig:res}
\end{figure}

\noindent\textbf{Pronunciation:} The Conformer model is widely used in speech recognition, so we adopted the pronunciation feature extractors based on the Conformer structure, as shown in Figure 2. First, we extract 80-dimensional log mel spectrograms from the raw audio. Then we use convolution downsampling in the time scale. The downsampled log mel spectrograms is then fed into the Conformer module, which follows the configuration in \cite{gulati2020conformer}. Then the predicted phoneme sequences are obtained through the CTC decoder and the attention decoder. The CTC decoder has a fully-connected layer. The attention decoder is location sensitive and has a decoder LSTM layer with a hidden size of 320. The training loss is a linear combination of the CTC and attention losses: 
\begin{equation}
	\mathcal{L} = \alpha \mathcal{L}_{CTC} + (1 - \alpha) \mathcal{L}_{ATT}
\end{equation}
Where $\mathcal{L}_{CTC}$ and $\mathcal{L}_{ATT}$ denote the loss of CTC and attention. $\alpha \in [0,1] $ is a hyper-parameter, and we set it to be $0.5$. After training, we keep the acoustic encoder part as a deeply embedded feature extractor. The pronunciation features encoded from raw audio are regarded as the pronunciation representations of the speech.

\noindent\textbf{Wav2vec:} Wav2vec 2.0 is a self-supervised speech representation learning method that can learn representations directly from raw audio signals without annotations. Its innovation lies in the use of the Transformer architecture, which can capture long-range dependencies. The model is trained using a masked contrastive predictive coding (CPC) objective to predict speech signals. The use of large-scale datasets and training processes improves the quality of the representations.

\subsection{Fusion Strategy for Features}
We use Transformer \cite{vaswani2017attention} to fuse the wav2vec feature with the other two features, respectively. We only use the encoder part of the Transformer. It is based on multi head attention mechanism. Multi head attention operates multiple self attention operations in parallel. The formula is as follows: 
\begin{equation}
	Attention(Q,K,V)= softmax(\frac{QK^T}{\sqrt{d_k}})V
\end{equation}
where $d_k$ is the key dimension. Here, we use the embedding of wav2vec as keys and values, and the embedding of duration and pronunciation as queries, respectively. The multi-head attention mechanism obtains $h$ different representations of (Q, K, V ), computes scaled dot-product attention for each representation, concatenates the results, and projects the concatenation through a feed-forward layer. It can be defined as:
\begin{equation}
	MultiHead(Q,K,V)= Concat(head_1,\ldots,head_h)W^O
\end{equation}
\begin{equation}
	head_i= Attention(Q{W_i}^Q,K{W_i}^K,V{W_i}^V)
\end{equation}

\subsection{Back-end Architecture}
Regarding the back-end architecture, we follow the conclusion in \cite{wang2021investigating} that a deep back-end is necessary when the front-end pre-training features are fixed. Our back-end architecture consists of a light convolutional neural network (LCNN) followed by two bi-directional recurrent layers with long short-term memory (LSTM) units, a global average pooling layer, and a fully connected (FC) output layer. We adopt the same configuration as presented in \cite{wang2021investigating}.

\section{Experiments}
\subsection{Dataset}
We employ five fake audio datasets. All of the models were trained on the ASVspoof2019\cite{todisco2019asvspoof} LA training sets. The ASVspoof2015\cite{wu2015asvspoof} is the most similar to the ASVspoof2019 LA for their audios are collected from the same datasets or conversion algorithms. The VCC2020\cite{yi2020voice} dataset is multilingual. The In-the-Wild\cite{muller2022does} dataset is collected from the real world. The ADD2022\cite{yi2022add} track2 dataset is Chinese and it is partially fake.

\begin{table}[t]
	
	%\vspace{-0.2cm}
	\caption{Statistics of experimental datasets.}
	%\vspace{-0.2cm}
	%\label{2013}
	
	\centering
	\begin{tabular}{c@{\hspace{0.1cm}}c@{\hspace{0.1cm}}c@{\hspace{0.1cm}}c }
		\toprule
		\textbf{Datasets} & \textbf{\#Real} & \textbf{\#Fake} & \textbf{Others}  \\
		\midrule
		ASVspoof2019(IN) & 7,355 & 63,882 & TTS and VC, English      \\
		\midrule
		ASVspoof2015(A) & 9,404 & 184,000 & TTS and VC, English     \\
		\midrule
		VCC2020(B) & 2,660 & 6,120 & VC, multilingual      \\
		\midrule
		In-the-Wild(C) & 18,863 & 11,816 & realistic, English     \\
		\midrule		
		ADD2022(D)  & 30,000  & 70,000 & partial fake, Chinese  \\
		
		\bottomrule
	\end{tabular}
	
\end{table}

\begin{table*}[t]
	
	%\vspace{-0.2cm}
	\caption{EER(\%) of our proposed different systems in in-domain and out-of-domain testing, where 'IN' denotes 'ASVspoof2019 LA' and ‘A’, ‘B’, ‘C’, ‘D’ denotes 'ASVspoof2015', 'VCC2020', 'in\_the\_wild' and 'ADD2020 track2', respectively. $O_1$ denotes concatenating and $O_2$ denotes the attention mechanism. When utilizing a single feature as input, neither concatenation nor attention fusion is employed. Results are the average obtained from three runs of each experiment with different random seeds.}
	%\vspace{-0.2cm}
	%\label{2013}
	
	\centering
	\begin{tabular}{c||cc|cc|cc|cc|cc }
		\toprule
		\multirow{2}{*}{\textbf{Feature}}   & \multicolumn{2}{c}{\textbf{IN}} &  \multicolumn{2}{c}{\textbf{A}} &  \multicolumn{2}{c}{\textbf{B}} &  \multicolumn{2}{c}{\textbf{C}} &  \multicolumn{2}{c}{\textbf{D}}\\
		\cline{2-11}
		& \textbf{$O_1$} & \textbf{$O_2$} & \textbf{$O_1$} & \textbf{$O_2$}& \textbf{$O_1$} & \textbf{$O_2$}& \textbf{$O_1$} & \textbf{$O_2$}& \textbf{$O_1$} & \textbf{$O_2$}  \\
		\hline
		Pron & 11.82&- & 25.31&- & 40.83 &-& 58.53&- &  45.38&-     \\
		Duration & 20.82 &- & 42.18 &-& 48.31 &-& 66.48&- &  48.51 &-    \\
	    Duration + Pron & 9.72& - & 32.74& -& 37.47& -& 59.43& -& 43.26& -\\
		\hline
		LFCC & 4.86&- & 28.15&- & 35.62&- & 62.53&- &  43.62&-     \\
		Pron + LFCC & 3.82 & 3.63 & 25.89 & 24.07 & 31.58 & 29.64 & 57.24& 56.77&37.86& 36.22\\
		Duration + LFCC    & 4.35 & 4.04  & 28.05 & 25.42 & 32.93 & 32.41& 60.08& 58.91&41.24& 39.79\\
		Duration + Pron +LFCC & 3.51 & \textbf{3.18}& 25.43 & \textbf{21.86} & 30.84 & \textbf{28.36}& 55.49& \textbf{54.22}&35.94& \textbf{34.82}\\
		\hline
		Wav2vec  & 3.16 &- & 6.59 &-& 19.33 &-& 43.81 &-& 35.19 &- \\
		Pron + Wav2vec & 2.83 & 1.97 & 4.85 & 3.28 & 17.28 & 15.31 & 40.79& 39.36&32.18 &31.52\\
		Duration + Wav2vec   & 3.08  &  2.44  &5.33  &4.25 & 18.05 & 17.54 & 42.15& 41.37&34.27 &32.68\\
		Duration + Pron + Wav2vec & 2.35 & \textbf{1.58} & 3.96 &\textbf{3.08} & 16.45 &\textbf{14.76} & 38.57& \textbf{36.84}& 30.77 &\textbf{29.53}\\ 
		
		\bottomrule
	\end{tabular}
	
\end{table*}

\subsection{Experimental Setup}
The Wav2vec XLSR model was obtained from the Fairseq project\footnote{ https://github.com/pytorch/fairseq/tree/main/examples/wav2vec/xlsr}. The model was pre-trained on a training set that includes Multilingual LibriSpeech, CommonVoice, and BABEL, which cover 8, 36, and 17 languages, respectively. This extensive training corpus allows the Wav2vec XLSR model to learn robust and diverse speech representations across a wide range of languages and accents. In order to train the Conformer model described in sections 2.1, we utilized the LibriSpeech dataset, which consists of 960 hours of audio recordings \cite{panayotov2015librispeech}. The corresponding text transcripts were used in conjunction with the LibriSpeech lexicon \footnote{http://www.openslr.org/11/} to obtain phoneme sequences for training the model.

The Wav2vec XLSR model has a dimension of 1024. In order to reduce the computational complexity, we apply a fully connected layer to reduce the dimensionality of the input features to 128. To form batches, we fix the length of each sample to 500 frames by truncating or concatenating. Thus, the shapes of the resulting wav2vec, duration and pronunciation features are $500\times128$, $500\times1$ and $500\times144$ , respectively. The audio sampling rate is 16k. For comparison, the baseline uses LFCC extracted with a frame length of 20 ms, a frame shift of 10 ms, and a 512-point fast Fourier transform (FFT). Each LFCC frame vector has a dimension of 60, including static, delta, and deltadelta components. In addition, 500 frames of the input is also needed at inference time.

For feature fusion, we use two methods: concatenation and the attention mechanism. For the former, we have directly concatenated the wav2vec features and the other two features from the temporal dimension, resulting in feature shapes of $500\times(128+1+144)$. For the latter, we first perform the attention transformation of wav2vec and the other two features separately, and then concatenate the obtained vectors to get the final representation. For the attention mechanism, we use 6 blocks and 8 heads.

To train the model, we use the Adam optimizer with a learning rate of $5 \times 10^{-5}$. The batch size is 32. The model is trained for 200 epochs. The EER \cite{cheng2004method} is used as the evaluation metric.

\section{Results and Discussion}
\subsection{Baseline}
The first few lines of Table 2 present the results of individual features on various datasets. It's worth noting that all models discussed in this paper were solely trained on the ASVspoof2019 LA dataset and then tested across different datasets. Based on our observations, testing across datasets is a challenging task. For instance, when evaluating LFCC, the EER on the ASVspoof2019 LA set was 4.86\%, but its performance declined to varying degrees on all four other datasets. In comparison, Wav2vec demonstrated better generalization than LFCC, with an EER of 6.59\% on the ASVspoof2015 LA test set, but performed poorly on VCC2020, in\_the\_wild, and ADD2022 datasets. This is because the generation of fake speech for different datasets and the recording environment of real speech vary. These findings support the motivation of our paper to enhance the generalization of detection models across datasets.

The second observation is that individual prosodic or pronunciation features exhibited poor performance on in-domain and out-of-domain tests. This could be attributed to the fact that prosodic features are one-dimensional, while wav2vec features are 128-dimensional and LFCC features are 60-dimensional. Due to their limited dimensionality, prosodic features contain less information compared to short-time spectral features and wav2vec features, which leads to the loss of acoustically relevant information in speech. In addition, the poor performance of pronunciation features could be due to the small size of the training data for the pronunciation extractor, which only contained 960 hours of data, compared to the 436k hours of training data for wav2vec.

\begin{table}[!t]
	\caption{Compare with the system of cross-dataset testing recently proposed. 'IN' denotes 'ASVspoof2019 LA', ‘A’ denotes 'ASVspoof2015', and ‘B’ denotes 'VCC2020'.}
	\centering
	\label{tab:la}
	\begin{tabular}{c||c|c|c}
		\hline
		\textbf{Methods}  & \textbf{IN} & \textbf{A} & \textbf{B}\\
		\hline
		Vanilla \cite{zhang2021empirical}&  2.29 & 26.30 & 41.66\\
		AUG \cite{zhang2021empirical}&  2.92 & 16.25 & 30.51\\
		MT-AUG \cite{zhang2021empirical}&  3.41 & 22.10 & 28.85\\
		ADV-AUG \cite{zhang2021empirical}&  3.23 & 14.38 & 27.07\\
		\hline
		Duration + Pron +LFCC (ours) & 3.18 & 21.86 & 28.36 \\
		Duration + Pron + Wav2vec (ours) & \textbf{1.58} & \textbf{3.08} & \textbf{14.76}\\
		\hline
	\end{tabular}
\end{table}

\subsection{Proposed Method Results}
The last few rows of Table 2 show the results of the proposed approach across datasets. We can draw the following conclusions: first, for both in-domain and out-of-domain tests, the performance of combining the prosodic features and pronunciation features with the wav2vec features is better than that of using the wav2vec features alone. This shows that using the prosodic and pronunciation features as auxiliary features has a positive impact. Specifically, concatenating the two features with the wav2vec features yielded the most noticeable performance improvement, with the EER decreasing from 6.59\% to 3.96\% on the ASVspoof2015 test set. This is due to the addition of wav2vec information from the prosodic and pronunciation dimensions. Second, the fusion method using the attention mechanism has better performance than the direct concatenating method. For example, when the front-end input is “duration + pron+ wav2vec”, the performance of the attention mechanism approach is 1.69\% better than the concatenate operation on the VCC2020 test set. This is because the attention mechanism allocates the weights of prosodic features and pronunciation features, which makes the model pay more attention to features that can distinguish real and fake audio, thus improving the generalization of the model. Third, in the cross-dataset test, the proposed method improves ASVspoof2015 more than the other three datasets. For example, when the front-end input is “duration + pron+ wav2vec”, the relative improvement of ASVspoof2015 with WAV2VEC features alone is 53.26\%, compared with 23.64\%, 15.90\%, and 15.76\% for VCC2020, in\_the\_wild, and ADD2022, respectively. This is because the dataset composition of ASVspoof2015 is similar to that of ASVspoof2019. Moreover, for LFCC, similar conclusions to wav2vec can be drawn. First, for both in-domain and out-of-domain tests, the performance of combining the prosodic features and pronunciation features with the LFCC features is better than that of using the LFCC features alone. Second, the fusion method using the attention mechanism has better performance than direct concatenating method. Third, in the cross-dataset test, the proposed method improves ASVspoof2015 more than the other three datasets.

Table 3 compares the proposed methods with the recently proposed cross-dataset testing systems \cite{zhang2021empirical}. When using “duration + pron + wav2vec” as the front-end feature, the system in this paper far outperforms other systems in both in-set and out-of-set performance. When using “duration + pron + LFCC” as the front-end feature, although the performance improvement is not as good as “duration + pron + wav2vec”, the performance is still competitive when considering both in-set and out-of-set results. In addition, we noticed a significant improvement in cross-dataset testing when using only wav2vec features. This is because wav2vec features are trained using only real speech data and have not been exposed to any fake audio, making them theoretically well-suited for generalizing to all types of fake audio. However, experimental results show that the generalization of wav2vec features is still influenced by the degree of match between the test and training sets. Our experimental results demonstrate that fusing prosodic and pronunciation features with wav2vec features can further improve the generalization of cross-dataset detection.

\section{Conclusion}
In this paper, we observe a significant performance degradation of existing fake audio detection systems in cross-dataset testing. This paper proposes multi-view features for fake audio detection, which attempts to capture more generalized features from the view of prosodic features, pronunciation features and wav2vec features. The results show that the prosodic and pronunciation features can be used as auxiliary features to improve the detection performance in and out of the domain. The fusion of the prosodic features and pronunciation features with wav2vec features is more effective by using the attention mechanism. In the future, we will explore different fusion strategies.
% THIS LIST IS PROVISIONAL - pending final version from Kate.
% \setlist{noitemsep,topsep=0pt,parsep=2pt,partopsep=0pt,leftmargin=1em}

\bibliographystyle{IEEEtran}
\bibliography{mybib}

% Generated by IEEEtran.bst, version: 1.13 (2008/09/30)
\begin{thebibliography}{10}
\providecommand{\url}[1]{#1}
\csname url@samestyle\endcsname
\providecommand{\newblock}{\relax}
\providecommand{\bibinfo}[2]{#2}
\providecommand{\BIBentrySTDinterwordspacing}{\spaceskip=0pt\relax}
\providecommand{\BIBentryALTinterwordstretchfactor}{4}
\providecommand{\BIBentryALTinterwordspacing}{\spaceskip=\fontdimen2\font plus
\BIBentryALTinterwordstretchfactor\fontdimen3\font minus
  \fontdimen4\font\relax}
\providecommand{\BIBforeignlanguage}[2]{{%
\expandafter\ifx\csname l@#1\endcsname\relax
\typeout{** WARNING: IEEEtran.bst: No hyphenation pattern has been}%
\typeout{** loaded for the language `#1'. Using the pattern for}%
\typeout{** the default language instead.}%
\else
\language=\csname l@#1\endcsname
\fi
#2}}
\providecommand{\BIBdecl}{\relax}
\BIBdecl

\bibitem{witkowski2017audio}
M.~Witkowski, S.~Kacprzak, P.~Zelasko, K.~Kowalczyk, and J.~Galka, ``Audio
  replay attack detection using high-frequency features.'' in
  \emph{Interspeech}, 2017, pp. 27--31.

\bibitem{font2017experimental}
R.~Font, J.~M. Esp{\'\i}n, and M.~J. Cano, ``Experimental analysis of features
  for replay attack detection-results on the asvspoof 2017 challenge.'' in
  \emph{Interspeech}, 2017, pp. 7--11.

\bibitem{novoselov2016stc}
S.~Novoselov, A.~Kozlov, G.~Lavrentyeva, K.~Simonchik, and V.~Shchemelinin,
  ``Stc anti-spoofing systems for the asvspoof 2015 challenge,'' in \emph{2016
  IEEE International Conference on Acoustics, Speech and Signal Processing
  (ICASSP)}.\hskip 1em plus 0.5em minus 0.4em\relax IEEE, 2016, pp. 5475--5479.

\bibitem{korshunov2016cross}
P.~Korshunov and S.~Marcel, ``Cross-database evaluation of audio-based spoofing
  detection systems,'' Tech. Rep., 2016.

\bibitem{das2019long}
R.~K. Das, J.~Yang, and H.~Li, ``Long range acoustic and deep features
  perspective on asvspoof 2019,'' in \emph{2019 IEEE Automatic Speech
  Recognition and Understanding Workshop (ASRU)}.\hskip 1em plus 0.5em minus
  0.4em\relax IEEE, 2019, pp. 1018--1025.

\bibitem{todisco2017constant}
M.~Todisco, H.~Delgado, and N.~Evans, ``Constant q cepstral coefficients: A
  spoofing countermeasure for automatic speaker verification,'' \emph{Computer
  Speech \& Language}, vol.~45, pp. 516--535, 2017.

\bibitem{sahidullah2015comparison}
M.~Sahidullah, T.~Kinnunen, and C.~Hanilçi, ``{A comparison of features for
  synthetic speech detection},'' in \emph{Proc. Interspeech 2015}, 2015, pp.
  2087--2091.

\bibitem{das2020assessing}
R.~K. Das, J.~Yang, and H.~Li, ``Assessing the scope of generalized
  countermeasures for anti-spoofing,'' in \emph{ICASSP 2020-2020 IEEE
  International Conference on Acoustics, Speech and Signal Processing
  (ICASSP)}.\hskip 1em plus 0.5em minus 0.4em\relax IEEE, 2020, pp. 6589--6593.

\bibitem{todisco2019asvspoof}
M.~Todisco, X.~Wang, V.~Vestman, M.~Sahidullah, H.~Delgado, A.~Nautsch,
  J.~Yamagishi, N.~Evans, T.~H. Kinnunen, and K.~A. Lee, ``{ASVspoof 2019:
  Future Horizons in Spoofed and Fake Audio Detection},'' in \emph{Proc.
  Interspeech 2019}, 2019, pp. 1008--1012.

\bibitem{tak2021end}
H.~Tak, J.~Patino, M.~Todisco, A.~Nautsch, N.~Evans, and A.~Larcher,
  ``End-to-end anti-spoofing with rawnet2,'' in \emph{ICASSP 2021-2021 IEEE
  International Conference on Acoustics, Speech and Signal Processing
  (ICASSP)}.\hskip 1em plus 0.5em minus 0.4em\relax IEEE, 2021, pp. 6369--6373.

\bibitem{xiao2015spoofing}
X.~Xiao, X.~Tian, S.~Du, H.~Xu, E.~Chng, and H.~Li, ``Spoofing speech detection
  using high dimensional magnitude and phase features: the ntu approach for
  asvspoof 2015 challenge.'' in \emph{Interspeech}, 2015, pp. 2052--2056.

\bibitem{attorresi2022combining}
L.~Attorresi, D.~Salvi, C.~Borrelli, P.~Bestagini, and S.~Tubaro, ``Combining
  automatic speaker verification and prosody analysis for synthetic speech
  detection,'' \emph{arXiv preprint arXiv:2210.17222}, 2022.

\bibitem{kamble2020advances}
M.~R. Kamble, H.~B. Sailor, H.~A. Patil, and H.~Li, ``Advances in
  anti-spoofing: from the perspective of asvspoof challenges,'' \emph{APSIPA
  Transactions on Signal and Information Processing}, vol.~9, p.~e2, 2020.

\bibitem{patil2017novel}
H.~A. Patil, M.~R. Kamble, T.~B. Patel, and M.~H. Soni, ``Novel variable length
  teager energy separation based instantaneous frequency features for replay
  detection.'' in \emph{INTERSPEECH}, 2017, pp. 12--16.

\bibitem{wu2015spoofing}
Z.~Wu, N.~Evans, T.~Kinnunen, J.~Yamagishi, F.~Alegre, and H.~Li, ``Spoofing
  and countermeasures for speaker verification: A survey,'' \emph{speech
  communication}, vol.~66, pp. 130--153, 2015.

\bibitem{patel2016effectiveness}
T.~B. Patel and H.~A. Patil, ``Effectiveness of fundamental frequency (f 0) and
  strength of excitation (soe) for spoofed speech detection,'' in \emph{2016
  IEEE International Conference on Acoustics, Speech and Signal Processing
  (ICASSP)}.\hskip 1em plus 0.5em minus 0.4em\relax IEEE, 2016, pp. 5105--5109.

\bibitem{pal2018synthetic}
M.~Pal, D.~Paul, and G.~Saha, ``Synthetic speech detection using fundamental
  frequency variation and spectral features,'' \emph{Computer Speech \&
  Language}, vol.~48, pp. 31--50, 2018.

\bibitem{xue2022audio}
J.~Xue, C.~Fan, Z.~Lv, J.~Tao, J.~Yi, C.~Zheng, Z.~Wen, M.~Yuan, and S.~Shao,
  ``Audio deepfake detection based on a combination of f0 information and real
  plus imaginary spectrogram features,'' in \emph{Proceedings of the 1st
  International Workshop on Deepfake Detection for Audio Multimedia}, 2022, pp.
  19--26.

\bibitem{lv2022fake}
Z.~Lv, S.~Zhang, K.~Tang, and P.~Hu, ``Fake audio detection based on
  unsupervised pretraining models,'' in \emph{ICASSP 2022-2022 IEEE
  International Conference on Acoustics, Speech and Signal Processing
  (ICASSP)}.\hskip 1em plus 0.5em minus 0.4em\relax IEEE, 2022, pp. 9231--9235.

\bibitem{wang2021investigating}
X.~Wang and J.~Yamagishi, ``Investigating self-supervised front ends for speech
  spoofing countermeasures,'' \emph{arXiv preprint arXiv:2111.07725}, 2021.

\bibitem{martin2022vicomtech}
J.~M. Mart{\'\i}n-Do{\~n}as and A.~{\'A}lvarez, ``The vicomtech audio deepfake
  detection system based on wav2vec2 for the 2022 add challenge,'' in
  \emph{ICASSP 2022-2022 IEEE International Conference on Acoustics, Speech and
  Signal Processing (ICASSP)}.\hskip 1em plus 0.5em minus 0.4em\relax IEEE,
  2022, pp. 9241--9245.

\bibitem{hsu2021hubert}
W.-N. Hsu, B.~Bolte, Y.-H.~H. Tsai, K.~Lakhotia, R.~Salakhutdinov, and
  A.~Mohamed, ``Hubert: Self-supervised speech representation learning by
  masked prediction of hidden units,'' \emph{IEEE/ACM Transactions on Audio,
  Speech, and Language Processing}, vol.~29, pp. 3451--3460, 2021.

\bibitem{gulati2020conformer}
A.~Gulati, J.~Qin, C.-C. Chiu, N.~Parmar, Y.~Zhang, J.~Yu, W.~Han, S.~Wang,
  Z.~Zhang, Y.~Wu \emph{et~al.}, ``Conformer: Convolution-augmented transformer
  for speech recognition,'' \emph{arXiv preprint arXiv:2005.08100}, 2020.

\bibitem{kharitonov2022textless}
E.~Kharitonov, J.~Copet, K.~Lakhotia, T.~A. Nguyen, P.~Tomasello, A.~Lee,
  A.~Elkahky, W.-N. Hsu, A.~Mohamed, E.~Dupoux \emph{et~al.}, ``textless-lib: a
  library for textless spoken language processing,'' \emph{arXiv preprint
  arXiv:2202.07359}, 2022.

\bibitem{vaswani2017attention}
A.~Vaswani, N.~Shazeer, N.~Parmar, J.~Uszkoreit, L.~Jones, A.~N. Gomez,
  {\L}.~Kaiser, and I.~Polosukhin, ``Attention is all you need,''
  \emph{Advances in neural information processing systems}, vol.~30, 2017.

\bibitem{wu2015asvspoof}
Z.~Wu, T.~Kinnunen, N.~Evans, J.~Yamagishi, C.~Hanil{\c{c}}i, M.~Sahidullah,
  and A.~Sizov, ``Asvspoof 2015: the first automatic speaker verification
  spoofing and countermeasures challenge,'' in \emph{Sixteenth annual
  conference of the international speech communication association}, 2015.

\bibitem{yi2020voice}
Z.~Yi, W.-C. Huang, X.~Tian, J.~Yamagishi, R.~K. Das, T.~Kinnunen, Z.-H. Ling,
  and T.~Toda, ``Voice conversion challenge 2020---intra-lingual semi-parallel
  and cross-lingual voice conversion,'' in \emph{Proc. Joint Workshop for the
  Blizzard Challenge and Voice Conversion Challenge 2020}, 2020, pp. 80--98.

\bibitem{muller2022does}
N.~M. M{\"u}ller, P.~Czempin, F.~Dieckmann, A.~Froghyar, and K.~B{\"o}ttinger,
  ``Does audio deepfake detection generalize?'' \emph{arXiv preprint
  arXiv:2203.16263}, 2022.

\bibitem{yi2022add}
J.~Yi, R.~Fu, J.~Tao, S.~Nie, H.~Ma, C.~Wang, T.~Wang, Z.~Tian, Y.~Bai, C.~Fan
  \emph{et~al.}, ``Add 2022: the first audio deep synthesis detection
  challenge,'' in \emph{ICASSP 2022-2022 IEEE International Conference on
  Acoustics, Speech and Signal Processing (ICASSP)}.\hskip 1em plus 0.5em minus
  0.4em\relax IEEE, 2022, pp. 9216--9220.

\bibitem{panayotov2015librispeech}
V.~Panayotov, G.~Chen, D.~Povey, and S.~Khudanpur, ``Librispeech: an asr corpus
  based on public domain audio books,'' in \emph{2015 IEEE international
  conference on acoustics, speech and signal processing (ICASSP)}.\hskip 1em
  plus 0.5em minus 0.4em\relax IEEE, 2015, pp. 5206--5210.

\bibitem{cheng2004method}
J.-M. Cheng and H.-C. Wang, ``A method of estimating the equal error rate for
  automatic speaker verification,'' in \emph{2004 International Symposium on
  Chinese Spoken Language Processing}.\hskip 1em plus 0.5em minus 0.4em\relax
  IEEE, 2004, pp. 285--288.

\bibitem{zhang2021empirical}
Y.~Zhang, G.~Zhu, F.~Jiang, and Z.~Duan, ``An empirical study on channel
  effects for synthetic voice spoofing countermeasure systems,'' \emph{Proc.
  Interspeech 2021}, pp. 4309--4313, 2021.

\end{thebibliography}

\end{document}